\documentclass[acmsmall]{acmart}

\usepackage{amsmath,amssymb,amsfonts}
\usepackage{algorithmic}
\usepackage{graphicx}
\usepackage{textcomp}
\usepackage{xcolor}
\usepackage{tabularx}
\usepackage{datatool}
\usepackage{graphicx}
\usepackage{subfig}
\usepackage[T1]{fontenc}
\usepackage[utf8]{inputenc}

\AtBeginDocument{%
  \providecommand\BibTeX{{%
    \normalfont B\kern-0.5em{\scshape i\kern-0.25em b}\kern-0.8em\TeX}}}

\acmConference[IT Research Symposium '20]{IT Research Symposium '20: School of Information Technology IT Research Symposium}{April 14, 2020}{Cincinnati, OH}
\acmBooktitle{IT Research Symposium '20: School of Information Technology IT Research Symposium,
  April 14, 2020, Cincinnati, OH}

\begin{document}

\title[The Relationship between Internet and Democracy]{Explaining the Relationship between Internet and Democracy in Partly Free Countries Using Machine Learning Models}

{
\author{Mustafa Sagir}
\affiliation{%
  \institution{Department of Political Science, University of Cincinnati}
  \streetaddress{2610 McMicken Circle}
  \city{Cincinnati, OH}
  \country{United States}}
\email{Anonymized@uc.edu}

\author{Said Varlioglu}
\affiliation{%
  \institution{School of Information Technology, University of Cincinnati}
  \streetaddress{2610 McMicken Circle}
  \city{Cincinnati, OH}
  \country{United States}}
\email{Anonymized@uc.edu}
}

\renewcommand{\shortauthors}{Sagir \& Varlioglu, 2020}

\begin{abstract}
\textbf{ABSTRACT}

Previous studies have offered a variety of explanations on the relationship between democracy and the internet. However, most of these studies concentrate on regions, specific states or authoritarian regimes. No study has investigated the influence of the internet in partly free countries defined by the Freedom House. Moreover, very little is known about the effects of online censorship on the development, stagnation, or decline of democracy. Drawing upon the International Telecommunication Union, Freedom House, and World Bank databases and using machine learning methods, this study sheds new light on the effects of the internet on democratization in partly free countries. The findings suggest that internet penetration and online censorship both have a negative impact on democracy scores and the internet's effect on democracy scores is conditioned by online censorship. Moreover, results from random forest suggest that online censorship is the most important variable followed by governance index and education on democracy scores. The comparison of the various machine learning models reveals that the best predicting model is the 175-tree random forest model which has 92\% accuracy. Also, this study might help "IT professionals" to see their important role not only in the technical fields but also in society in terms of democratization and how close IT is to social sciences. 

\end{abstract}



\maketitle

\section{Introduction}

In a speech at the Paul H. Nitze School of Advanced International Studies of the Johns Hopkins University, March 9, 2000, on China W.T.O. agreement, President Clinton said: “In the new century, liberty will spread by cell phone and cable modem. Now there is no question China has been trying to crack down on the Internet. Good luck! That's sort of like trying to nail Jell-O to the wall" \cite{clinton2000speech}. This perspective shows the clear optimistic view of the internet's prospective influence on democratization in the early 2000s. However, leaders are more cautious about this optimistic prospect with the different developments in the following years. First of all, internet penetration has kept on increasing in every country in the world. Especially, with the rise of smartphones, more people have access to the internet and people spend more time on the internet. International Telecommunication Unit reports that in 2019, 53.6 percent of the global population, are using the Internet . This percentage was only 16.8 percent in 2005 \cite{itu2019}. Also, social media usage has significantly increased. For instance, 85 percent of the adult population uses the Internet regularly and 80 percent of those people also use Facebook in the US \cite{greenwood2016social}. The use of the internet and social media in the protests around the world strengthened the arguments that the internet can help democratization by providing an environment for sharing information and organizing.


On the other hand, Freedom House reported in 2019 that democracy scores have kept declining since 2005 \cite{FreedomHouse2019}. This 13 year of decrease is not limited to some regions or only some categories. Countries from free, partly free, and not free countries and countries from all regions have lower scores in 2018 than in 2005 \cite{FreedomHouse}. Thus, after the third wave of democratization which started with the Carnation Revolution in Portugal in 1974 and accelerated after the end of the Cold War, a backward course has been going on.


After the optimistic statements of the western leaders, we have seen that authoritarian regimes have used the internet to track the political activist, prosecute the dissidents, and spread their propaganda. Huntington argues “democracies are created not by causes but by causers” \cite{huntington1968political}. Drawing upon the same analogy, autocracies are also created by causers and according to one point of view, the internet gave more opportunities for authoritarian leaders to protect and establish their regimes. For example, digital rights group Access Now’s 2018 report reveals that in 2018 the internet is shut down 196 times at a national or regional level more than 90 times in 2017 \cite{acccessnow}.

Moreover, the internet as a double-edged sword can help share news and information but it can also be used to spread inaccurate information, manipulate voters, increase the polarization and hatred toward other people. 

A variety of studies investigated the relationship between the internet and democracy. Some studies found a positive association, but some studies which especially focus on case studies in authoritarian regimes found a negative effect of the internet on democratization. Also, some studies explored the effects of diffusion of false news through the internet and especially via social media. However, most of these studies concentrate on regions, specific states or authoritarian regimes. Freedom House puts countries in three categories; free, partly free, and not free. According to Freedom House 2019 Report \cite{FreedomHouse}, 63 out of 205 countries are in the partly free category in 2019, but no study has explored the impact of the internet in these countries. Moreover, Freedom House prepares the "Censorship on Net" reports since 2013 \cite{freedomonthenet}, but no study has investigated the relationship between internet access, online censorship, and democratization. 

The purpose of this quantitative study is to explore the influence of internet penetration and online censorship on democratization controlling economic, educational, life quality, and state capacity variables. Drawing upon International Telecommunication Union (ITU) internet penetration data \cite{itu2019}, Freedom House Democracy and internet censorship scores \cite{FreedomHouse}, World Bank World Development Index \cite{Worldbank2019}, and World Bank Governance Index data for 2013-2017 \cite{WorldbankGovernance}, this study sheds light on the effects of the internet and online censorship on democratization in partly free countries. The current study applies multivariate regression analysis and machine learning models to better understand the importance of variables affecting the democracy scores. Thus, there are several key areas where this study makes an original contribution to this growing area of research. 	

\section{Theoretical Perspectives and Literature}

There is a large volume of published studies describing the role of several factors in promoting, accelerating or preserving democratization. Moreover, in parallel with the people's access to the internet around the world, a large and growing body of literature has investigated the relationship between the internet and democracy. These studies can be broadly categorized under three distinct groupings. The first group of studies argues that with free access to information, knowledge sharing without any constraint, and the spread of political knowledge, the internet changes people’s political attitudes and they found a positive effect of the internet on democratization \cite{pirannejad2017can, stoycheff2016differential, breuer2014online, bailard2012field, lei2011political, settle2016posting, massoud2019protests, zang2019reversing}.

The second group of studies that especially focus on authoritarian regimes argues that the internet gives a distinct means to track the political activists, prosecute the dissidents, and spread the regime propaganda, which in turn limits the democratization \cite{groshek2017time, rod2015empowering, kalathil2001internet, griffiths2019great, zhong2017does}.

The third group of studies explored the effects of diffusion of false news through the internet and especially via social media. These studies argue that the internet and especially social media is not beneficial but detrimental to democracy \cite{bartlett2018people, deb2017social, mcchesney2016rich, howard2016political,bradshaw2018challenging,deibert2019road}

Although an extensive body of research has been carried out and studies over the past two decades have provided valuable information on the relationship between internet/social media use and several aspects of democracy, no single study exists which specifically focuses on partly free countries. Moreover, little is known about the effects of online censorship on the development, stagnation, or decline of democracy.

\section{The Current Study}

This quantitative study tries to fill these gaps by analyzing the relationship between internet penetration and democracy in partly free countries. This study also investigates the interaction between internet penetration and online censorship to see whether the democracy scores of partly free countries are conditioned by the level of online censorship that exists in those states.

\subsection{Research Question and Hypotheses}

The primary aim of the study is to examine the effects of internet penetration and online censorship on democratization in partly free countries controlling the economic, educational, and life quality variables. Thus, the central \textbf{research question} is:  “To what extent does internet promote, consolidate, and protect democracy in partly free countries?”

As outlined in the literature review section, there are contradicting arguments on the effects of the internet on democratization. Consequently, there are competing hypotheses and expectations about how the increase in the usage of the internet may affect democratization. 

Drawing on the argument that the internet accelerates the democratization, \textbf{Hypothesis 1}: An increase in the accessibility of the internet increases the democracy scores in partly free countries. 

The second argument is that governments can use censorship tactics to prevent information sharing, track dissidents, and block social unrest by using the internet, which in turn decline democratization. Hence, \textbf{Hypothesis 2}: An increase in internet censorship prevents democratization and decrease democracy scores. 

In the third model, we investigate the moderation effect of online censorship. Accordingly, \textbf{Hypothesis 3}: Internet penetration’s effect on democratization is conditioned by online censorship.

\subsection{Dependent Variable}

The dependent variable in the current study is "Freedom House Total Aggregate Scores" for the period 2013-2017. Freedom House assesses Political Rights in 10 indicators and Civil Liberties in 15 indicators. Each indicator has calculated from 0-4. Thus, each country or territory has a Political Right score of 0-40, Civil Liberties score of 0-60, and Total Aggregate scores of 0-100. This study uses the Freedom House Total Aggregate scores as the dependent variable.

\subsection{Independent Variables}

The primary predictors of interest in the study are internet penetration and online censorship. Internet penetration is measured by the share of the population using the internet and obtained from the ITU website \cite{itu2019}. Online censorship is measured by Freedom House \cite{FreedomHouse} with a methodology that includes 21 areas in three categories. These categories are Obstacles to Access, Limits on Content, and Violations of User Rights. Freedom on Net scores is the measure of how governments and non-state actors around the world limit the fundamental rights online. Each country has a score of 0-100, 0 being the most restrictive and 100 being the most free.

\subsection{Control Variables}

There is a vast literature attempting to explain the variables influencing the emergence and the survivability of democracies. Drawing on these previous studies, this study employs variables from the World Bank World Development Indicators Database and Worldwide Governance Indicators Database. The variables related to economy, education, life quality, urbanization, and demographics were obtained from World Bank World Development Indicators Database These are; GDP per Capita, Annual percentage of GDP per Capita Growth, Unemployment percentage of total population, Foreign Direct Investment percentage of GDP, Inflation, Government Expenditure on Education percentage of GDP, Military Expenditure percentage of GDP, Life Expectancy, Number of Infant Deaths, Urban Population, and Annual percentage of Urban Population Growth. Since there are no consistent data on Educational Attainment, Income Distribution, and Share of the Middle Class, these variables are omitted from the model. 

We also included the variables related to state institutional capacity from five of the six World Bank Worldwide Governance Index elements of governance; Political Stability and Absence of Violence, Government Effectiveness, Regulatory Quality, Rule of Law, and Control of Corruption. The scores for these variables are range from -2.5 to 2.5, with greater values equivalent to better governance.

\subsection{Analytic Approach}

To be able to assess the effects of internet penetration and online censorship on democracy scores, first, we looked at the descriptive analysis to see how the variables are distributed across countries. Second, we investigated the bivariate analysis to see the correlation between variables. After finding that some variables are highly correlated with each other, we created a state capacity index which is composed of five Worldwide Governance Index’s five elements of governance. As a next step, we moved to multivariate analysis to be able to analyze the relationships among all variables to get a true picture. We scaled all the independent variables and ran the OLS Regression. Then, we created an interaction term between internet penetration and online censorship to see if the internet’s effect on democracy scores is conditioned by the level of online censorship in the countries in the sample and ran the OLS Regression again. After the regression analysis, we conducted decision tree, SVR, and random forest machine learning models analysis to find which variables are more important in the causality for democracy scores. We compared the test set results with actual results. Random Forest model has higher accuracy in predicting the democracy scores.

In addition, we conducted diagnostic tests to ensure that assumptions are met. We checked the multicollinearity and outliers in the data by careful selection of predictors and combining related measures into scales so that the collinearity issue was reduced and did not seem to be a problem.

\section{Findings}
\subsection{Descriptive Findings}

For the 5-year period during 2013-2017, Freedom House Aggregate scores range from 25 to 79 in the countries which were in defined as partly free countries. The internet penetration levels and online censorship scores are also greatly varied and are ranging from 5.05-84.45 for the internet penetration and 24-71 for the censorship scores. Venezuela and Turkey are the countries that showed high levels of decline in democracy scores and an increase in online censorship scores while Tunisia is the country that showed improvements in many areas. 

\subsection{Bivariate Relationship}

The bivariate correlations shown in Table 1 demonstrate the baseline associations between internet penetration, online censorship, various development, and governance index variables and the dependent variable Freedom House Aggregate Scores. Internet penetration and FH scores have a positive correlation, but it is not statistically significant. However, there is a statistically significant negative correlation between online censorship and FH scores. Moreover, there is a statistically significant positive correlation between democracy scores and GDP per capita growth, Foreign Direct Investment, and Governance Index.

\begin{tiny}

\begin{table}
  \caption{Correlation Table}
  \label{tab:table1}
  \begin{tabular}{lcccccccccccccccc}
    \toprule
\textbf{Variables}&	1&	2&	3&	4&	5&	6&	7&	8&	9&	10&	11&	12&	13&	14&	15&	16\\
    \midrule

\textbf{Year}&	1\\															
\textbf{Internet}& 	0.24*&	1\\														
\textbf{Censorship}&	0.06&	-0.12&	1	\\												
\textbf{GDP}&	0.03&	0.57*&	0.02&	1	\\											
\textbf{GDP} \textbf{Grow}.&	-0.05&	-0.21*&	-0.14&	0&	1	\\										
\textbf{Unemp}.&	0.05&	0.21*&	-0.2*&	-0.16&	-0.08&	1\\										
\textbf{FDI}&	-0.01&	0.33*&	-0.23*&	0.72*&	0.06&	0&	1\\									
\textbf{Inflation}&	0.12&	0.12&	0.25*&	0.06&	-0.41*&	0.01&	-0.14&	1\\								
\textbf{Mil} \textbf{Exp}.&	0	&0.4*&	0.08&	0.13&	-0.2*&	0.36*&	0.2*&	-0.25*&	1\\							
\textbf{Educ}.&	0.01&	0.05&	-0.19*	&-0.14&	-0.16&	0.15&	-0.1&	0.03&	-0.19*&	1	\\					
\textbf{Life} \textbf{Exp}.&	0.07&	0.71*&	0.05&	0.51*&	0.05&	0.15&	0.29*&	-0.03&	0.45*&	-0.1&	1\\					
\textbf{Pop}.& 15-64&	0.03&	-0.34*&	0.25*&	-0.16&	0.27*&	-0.38*&	-0.33*&	-0.05&	-0.37*	&-0.23*&	-0.21*&	1\\				
\textbf{Urb}. \textbf{Pop}.&	0.06&	0.87*&	-0.03&	0.59*&	-0.33*	&0.2*&	0.31*&	0.17&	0.4*&	0.03&	0.64*&	-0.17&	1	\\		
\textbf{Female}&	0&	-0.11&	-0.27*&	-0.39*&	-0.08&	0.39*&	-0.23*&	0.03&	0.1&	0.24*&	-0.11&	-0.32*&	-0.19*&	1	\\	
\textbf{Gov}.\textbf{Ind}.&	0&	0.45*&	-0.28*&	0.7*&	0.29*&	0.05&	0.73*&	-0.33*&	0.2*&	-0.18*&	0.54*&	-0.25*&	0.39*&	-0.25*&	1\\	
\textbf{FH} \textbf{Scores}&	0.03&	0.02&	-0.52*&	0.03&	0.17*&	-0.02&	0.09*&	-0.21*&	-0.21*&	0.07&	0.17&	0.15&	0.07&	0.11&	0.3*&	1\\

  \bottomrule
  \end{tabular}
    \begin{tabular}{l}
*. Correlation is significant at the 0.05 level, 2-tailed)\\
\end{tabular}
\end{table}

\end{tiny}

\subsection{Multivariate Relationship}

As a next step, we moved to multivariate analysis to be able to get a true picture. Table 2 shows the results of regression models. In the first model, just internet penetration was applied as an independent variable. In the second model, censorship scores are put in the model and internet penetration was removed from the model. In the third model, both variables are included and in the fourth model, the interaction term between internet penetration and online censorship was included to see whether internet penetration’s effect on Freedom House scores is conditioned by online censorship. 


\begin{tiny}

\begin{table}
  \caption{Multivariate Analysis}
  \label{tab:table2}
  \begin{tabular}{lcccc}
    \toprule
VARIABLES&	FH Scores (1)&	FH Scores (2)&	FH Scores (3)&	FH Scores (4)\\
    \midrule

\textbf{Year}&	1.35&	0.42&	2.03**&	2.06**\\
	&(0.97)&	(0.79)&	(0.86)&	(0.84)\\
\textbf{Internet} \textbf{Penetration}&	-5.45**	&	-8.61***&	-7.94***\\
&	(2.52)	&	(2.28)&	(2.25)\\
\textbf{Online} \textbf{Censorship}	&	-5.17***&	-6.13***&	-6.62***\\
	&	(1.09)&	(1.06)&	(1.05)\\
c.\textbf{InternetPenetration}\#c.\textbf{OnlineCensorship}&&&&				-3.32**\\
				&&&&(1.33)\\
\textbf{GDP} \textbf{Per} \textbf{Capita}&	-5.84***&	-2.47&	-2.88&	-1.30\\
&	(2.04)&	(1.97)&	(1.87)&	(1.93)\\
\textbf{GDP} \textbf{Per} \textbf{Capita} \textbf{Growth}&	-1.39&	-2.24*&	-1.73&	-1.58\\\\
	&(1.27)&	(1.18)&	(1.12)	&(1.10)\\
\textbf{Inflation}&	-2.50*&	-1.18&	-1.56&	-0.46\\
&	(1.44)&	(1.35)&	(1.28)&	(1.33)\\\\
\textbf{Military} \textbf{Expenditure} (\% GDP)&	-5.62***&	-2.94**	&-3.96***&	-4.77***\\
	&(1.50)	&(1.40)&	(1.35)&	(1.36)\\
\textbf{Government} \textbf{Expenditure} \textbf{on} \textbf{Education}	&0.74&	0.57&	0.06&	0.59\\
&	(1.12)&	(1.04)	&(0.99)	&(0.99)\\
\textbf{Life} \textbf{Expectancy}&	6.24***&	4.33**&	6.71***	&6.63***\\
&	(2.09)&	(1.83)&	(1.85)&	(1.80)\\
\textbf{Number} \textbf{of} \textbf{Infant} \textbf{Deaths}	&2.69&	-0.16&	1.28&	0.45\\
&	(1.75)&	(1.59)&	(1.56)&	(1.56)\\
\textbf{Population} \textbf{Ages15}-64&	0.38&	4.22***	&1.53&	1.28\\
&	(1.71)	&(1.42)&	(1.52)&	(1.49)\\
\textbf{Population} \textbf{Growth}&	-4.07&	-2.89&	-0.89&	-1.60\\\\
	&(2.69)&	(2.51)&	(2.43)	&(2.39)\\
\textbf{Urban} \textbf{Population}&	6.89***	&0.30&	6.51***&6.59***\\
&	(2.50)&	(1.55)&	(2.21)&	(2.16)\\
\textbf{Urban} \textbf{Population} \textbf{Growth}&	4.32&	3.36&	0.28&	3.17\\
	&(2.75)	&(2.52)	&(2.52)	&(2.72)\\
\textbf{Female} \textbf{Population}&	4.39***&	2.37&	1.93&	2.49*\\
&	(1.57)&	(1.53)&	(1.45)&	(1.43)\\
\textbf{Governance} \textbf{Index}&	5.59***&3.80**&	3.69**&	3.17*\\
&(1.94)&	(1.84)&	(1.74)&	(1.71)\\
\textbf{Constant}&	50.73***&	50.73***&	50.73***&	50.32***\\
&	(0.83)&	(0.77)&	(0.73)&	(0.73)\\

  \bottomrule
\end{tabular}

\begin{tabular}{l}
Standard errors in parentheses, *** p<0.01, ** p<.05, * p<0.1\\
\end{tabular}

\end{table}

\end{tiny}

According to regression analysis, both internet penetration and online censorship have a negative association with democracy scores. Moreover, the interaction term in the fourth model shows that the internet's effect on democracy scores is conditioned by online censorship.


\subsection{Decision Tree and Random Forest}

After the regression analysis, we conducted decision tree, SVR, and random forest machine learning models analysis to find which model has higher accuracy in predicting the democracy scores and which variables are more important in the causality for democracy scores. Then, we compared the test set results with actual results. Thus, we started with the decision tree. Figure 1 shows the decision tree. As seen in the picture, the machine first goes to censorship, then GDP growth and population, then life expectancy, foreign direct investment, and urban population growth according to the results of the previous inquiry.

\begin{figure}[h]
  \centering
  \includegraphics[width=0.7\columnwidth]{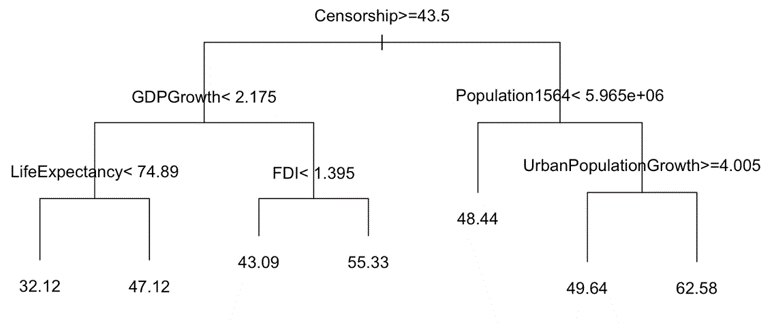}
  \caption{Decision Tree}
  \Description{Decision Tree}
\end{figure}

Then we moved to random forest. After trials of a different number of trees in Random Forest, the 175-tree model gave greater accuracy in predicting the Freedom House aggregate scores. We compared the prediction results in each of the models,; SVR, decision tree and various combinations of random forests. Random Forest with 175 tree gave the best prediction accuracy. We compared the predictions of each model. According to our comparison, the prediction accuracy of the regression and the decision tree models are close, 84.6 percent and 84.9, respectively. 500-tree random forest's prediction accuracy is 87.2 percent and 175-tree random forest's prediction accuracy is 91.9 percent.
Furthermore, Figure 2 shows the dot chart of the importance of variables based on a 175-tree random forest. As seen in the figure, online censorship is the most important variable in the model whereas internet penetration is not a very significant variable. 

\begin{figure}[h]
  \centering
  \includegraphics[width=0.8\columnwidth]{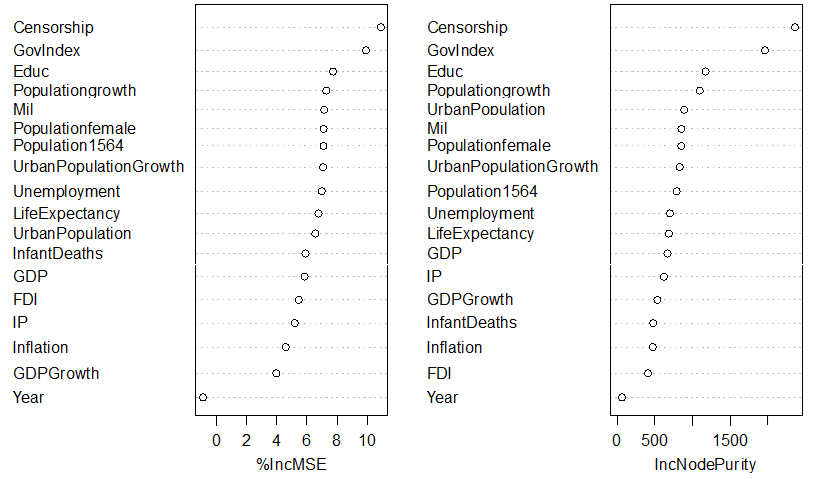}
  \caption{Dot chart of variable importance scores based on Random Forest with 175 trees.}
  \Description{Random Forest}
\end{figure}

Overall, these results indicate that there is a strong negative impact of online censorship on democracy scores whereas internet penetration’s influence is more mixed. Also, consistent with the earlier studies and theories, state institutional capacity and urban population have a positive association while military expenditure percentage of GDP has a negative association with democratization. Moreover, 175-tree random forest models' predictions are more accurate than those of a decision tree or regression models. Together these results provide important insights into internet-democracy relationship literature.

\subsection{Discussion and Conclusion}

This study certainly adds to our understanding of the internet democracy relationship and provides a different perspective. The current study aimed to determine the effects of internet penetration and online censorship on democracy scores in partly free countries. The most obvious finding to emerge from this study is that online censorship has a strong negative impact on democratization in the studied countries (\textbf{Hypothesis 1}). However, we are cautious to have a casual argument here because we could not determine which direction the causality runs. It is also possible that countries use online censorship as a tool when they become more anti-democratic. For instance, Hellmeier claims that monarchies, regimes with greater levels of protests, and less opposition in the political field are more likely to censor the Internet\cite{hellmeier2016dictator}. 

In addition to the online censorship’s impact, internet penetration, overall, has a negative effect on democratization (\textbf{Hypothesis 2}), however, the partial dependence plots in Figure 3 after the random forest analysis reveal that the internet's effect is more complicated and similar to a reverse U shape. Another significant finding to emerge from this study is that the internet's effect on democratization is conditioned by online censorship (\textbf{Hypothesis 3}). Findings also suggest that state education, and urban population, and a state's institutional capacity as measured by governance index are among the variables which have a positive association with democracy scores. On the other hand, an increase in the military expenditure percentage of GDP decreases the democratization. 

Moreover, with the machine learning models, by using the independent and control variables, we can predict the democracy scores of partly free countries with a prediction accuracy between 84 and 92 percent. On average, random forest with 175-tree gave the best prediction accuracy with 92 percent accuracy.  

\begin{figure}[h]
  \centering
  \includegraphics[width=1.0\columnwidth]{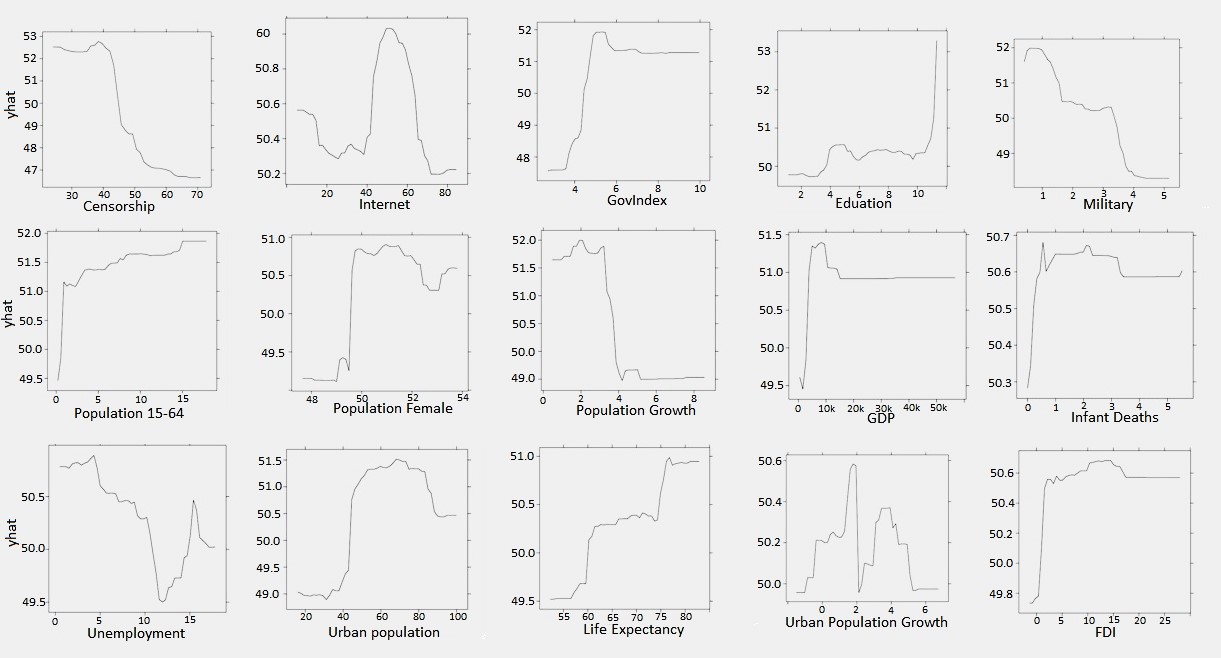}
  \caption{The Partial Dependence Plots}
  \Description{The Partial Dependence Plots}
\end{figure}

These findings have significant implications for the understanding of how online censorship influence the democratization in partly free countries. The present study is one of the first attempts to explore this relationship. The findings of the study will lay the groundwork for future research and help to improve explanations of the internet democracy relationship studies. Further research, especially on the online censorship-democracy relationship, would help us to establish a higher degree of accuracy on this issue.

\bibliographystyle{ACM-Reference-Format}
\bibliography{references}

\end{document}